\input{aipcheck}

\documentclass[
    ,final            
  ]
  {aipproc}

\layoutstyle{8x11single}

\newcommand{\gray}[1]{$\gamma$-ray{#1}}
\newcommand{\grays}[1]{$\gamma$-rays{#1}}

\newcommand{\pubjournal}[6] {#1, #2 {\bf #3}, #4 (#5).}

\newcommand{\icrc}{{\it Int. Cosmic Ray Conf.}}

\newcommand{\aap}{{\it Astron. Astrophys.}}
\newcommand{\apj}{{\it ApJ}}

\begin{document}

\title{Very High Energy Gamma Rays from Supernova Remnants and Constraints on 
the Galactic Interstellar Radiation Field}


\author{Troy A. Porter}{
  address={Santa Cruz Institute for Particle Physics, 
  University of California, Santa Cruz, CA 95064},
  email={tporter@scipp.ucsc.edu},
  thanks={Supported in part the US Department of Energy}
}

\author{Igor V. Moskalenko}{
  address={Hansen Experimental Physics Laboratory,
  Stanford University, Stanford, CA 94305},
  email={imos@stanford.edu},
  thanks={Supported in part by NASA APRA grant},
  altaddress={Kavli Institute for Particle Astrophysics and Cosmology,
    Stanford University, Stanford, CA 94309}
}

\author{Andrew W. Strong}{
  address={Max-Plank-Institut f\"ur extraterrestrische Physik, 
  Postfach 1312, D-85741 Garching, Germany},
  email={aws@mpe.mpg.de}
}


\begin{abstract}
The large-scale Galactic interstellar radiation field (ISRF) is the result
of stellar emission and dust re-processing of starlight.
Where the energy density of the ISRF is high (e.g., the Galactic Centre), 
the dominant \gray\ emission in individual supernova remnants (SNRs), 
such as G0.9+0.1, 
may come from inverse Compton (IC) scattering of the ISRF.
Several models of the ISRF exist.
The most recent one, which has been calculated by us, predicts a significantly
higher ISRF
than the well-used model of Mathis, Mezger, and Panagia \cite{MMP1983}.
However, comparison with data is limited to local observations.
Based on our current estimate of the ISRF we predict the gamma-ray 
emission 
in the SNRs G0.9+0.1 
and RXJ1713, and pair-production absorption features above 20 TeV in the 
spectra of G0.9+0.1, J1713-381, and J1634-472.
We discuss how GLAST, along with current and future very high energy 
instruments, may be able to provide upper bounds on the 
large-scale ISRF.
\end{abstract}

\maketitle


\section{Interstellar Radiation Field}
The ISRF calculation uses a model for the
distribution of stars in the Galaxy, a model for the dust distribution and
properties, and a treatment of scattering, absorption, and subsequent
re-emission of the stellar light by the dust.
A brief description of our calculation is available in 
\cite{PS2005,MPS2006,PMS2006}; full details will be available in a 
forthcoming publication \cite{PSD2007}.

\section{Gamma-ray Emission from Supernova Remnants and Attenuation on the ISRF}
\begin{figure}[htb]
\includegraphics[height=.28\textheight]{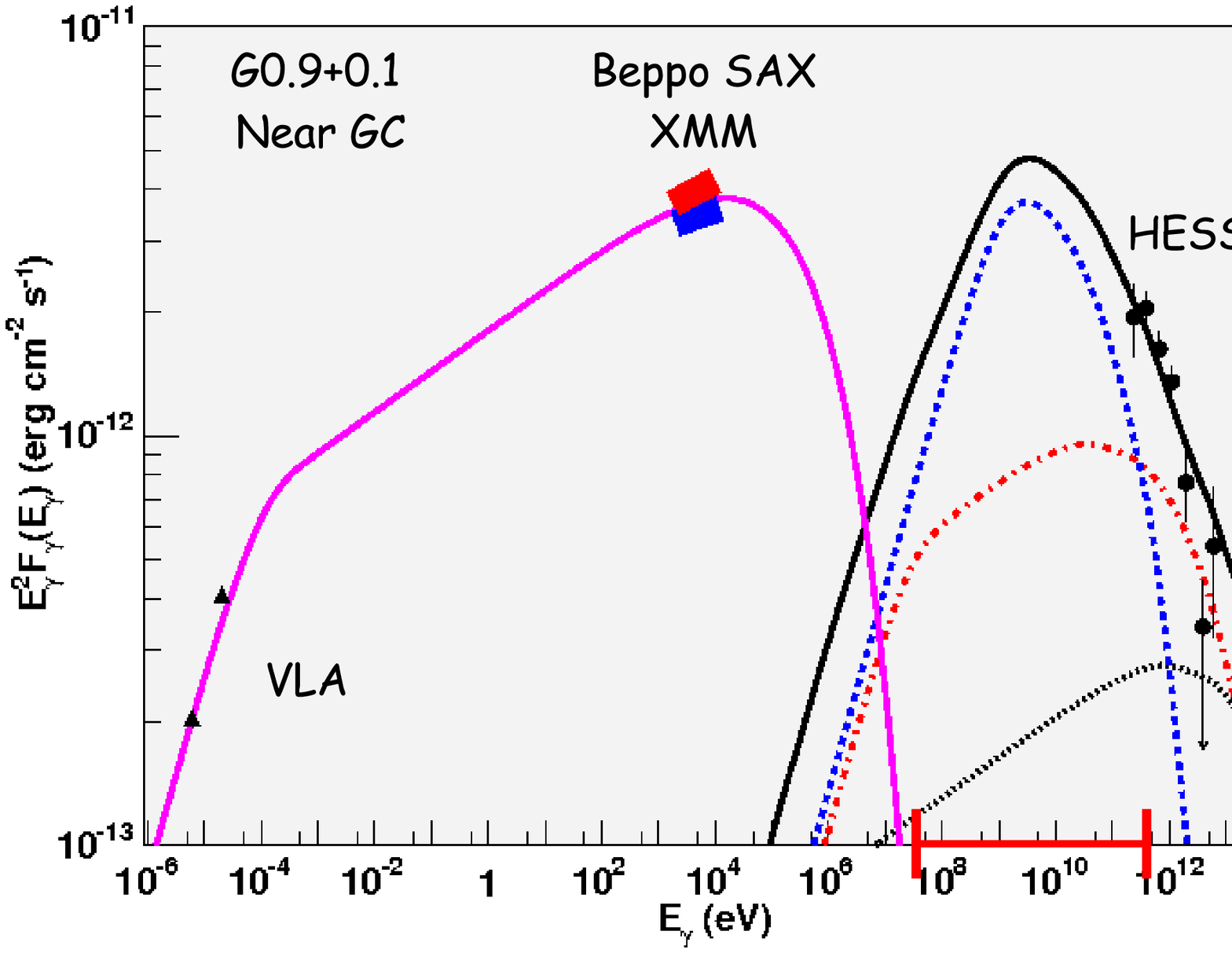}
\includegraphics[height=.28\textheight]{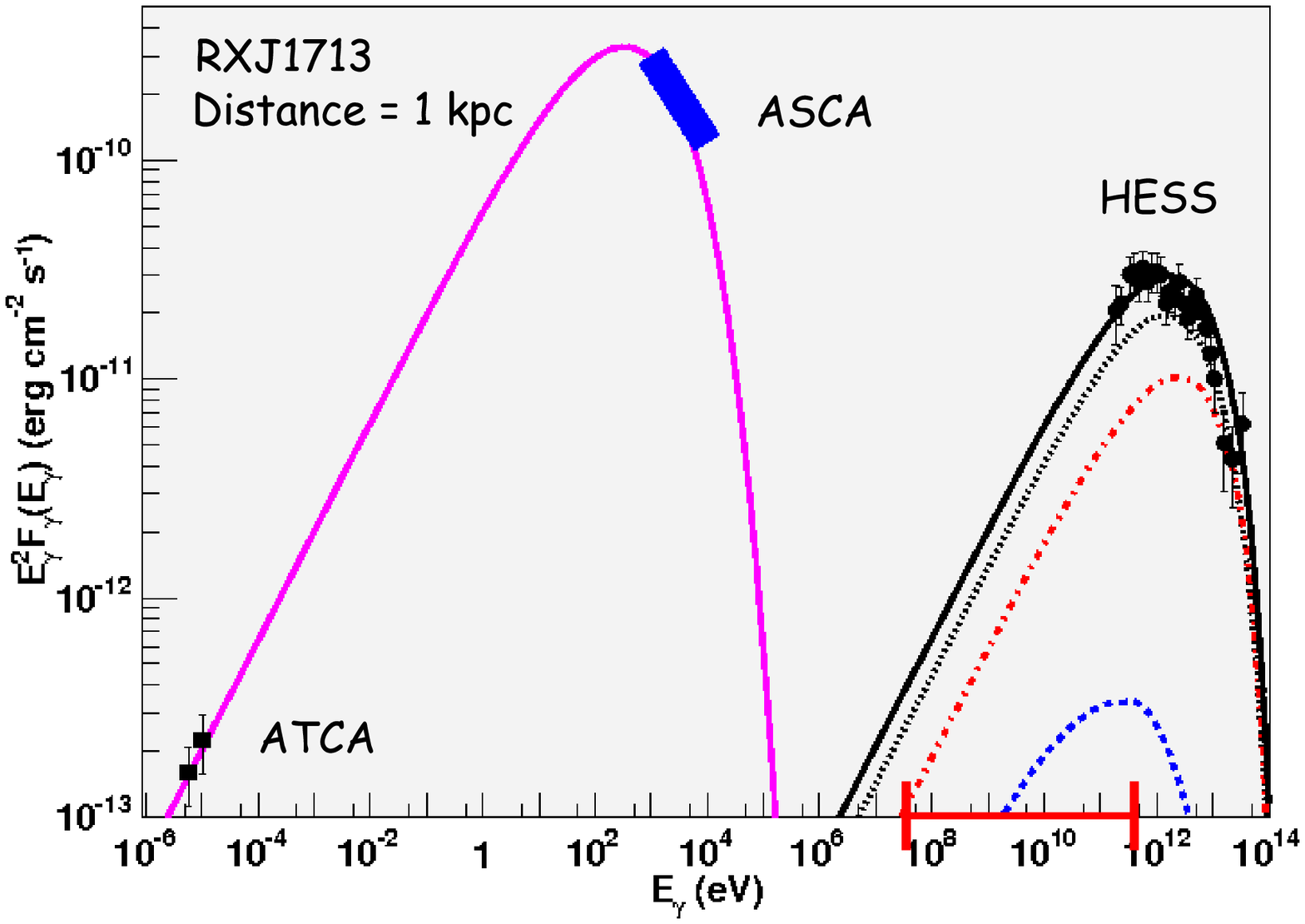}
\caption{{\it Left:} Flux spectrum of G0.9+0.1. 
Line-styles: solid, total synchrotron and IC flux; dashed, optical IC;
dash-dotted, infra-red IC; dotted, CMB IC. 
The GLAST energy range is 
indicated by the barred red line. 
{\it Right:} Flux spectrum of RXJ1713. Line-styles as for left panel. 
Data are summarised in \cite{PMS2006}.}
\label{fig1}
\end{figure}

The evidence for particle
acceleration in supernova shells comes from electrons whose synchrotron
emission is observed in radio and X-rays.
Recent observations by the HESS instrument
reveal that supernova remnants (SNRs) also emit TeV \grays.
We have considered one-zone leptonic models and fitted the multi-wavelength
spectra of the supernova remnants (SNRs) G0.9+0.1 and RXJ1713 
observed by the HESS \cite{PMS2006}.
Figure~\ref{fig1} shows our results.
For the inner Galaxy, the emission in the GLAST energy range is dominated
by IC scattering on the optical and infrared components of the ISRF, while 
scattering on the CMB dominates the emission toward the outer Galaxy.

\begin{figure}
\includegraphics[height=.22\textheight]{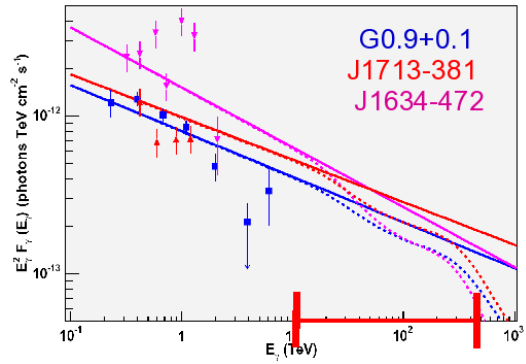}
\caption{Attenuation on the ISRF for in-plane HESS sources.
Line-styles: solid line, intrinsic spectrum; dashed line, attenuated spectrum.
Data and locations of sources are summarised in \cite{HESS2006b}.
The energy range of future experiments, such as TenTen \cite{Rowell2005} is 
indicated by the barred red line.} 
\label{fig2}
\end{figure}

We have made a calculation of 
attenuation of very high energy (VHE) \grays\ in the Galaxy 
using the new ISRF which
takes into account its nonuniform spatial and angular
distributions \cite{MPS2006}.
Figure~\ref{fig2} shows the effect of the attenuation on the ISRF 
for several HESS sources located at different positions in the Galactic plane.
The majority of the attenuation occurs on the infrared component of the ISRF
\cite{MPS2006}.

\section{Conclusions}
The IC emission of individual SNRs can include significant contributions
by the ISRF.
In the outer Galaxy the emission is predominantly from IC scattering
on the cosmic microwave background (CMB).
Thus, observations of outer Galaxy SNRs could
help determine intrinsic properties for the studied source class.
In turn, these could be applied to similar objects in the inner Galaxy.
By assuming that the ISRF is the sole target photon field, observations 
of inner Galaxy SNRs may provide an upper limit on the optical component 
of the ISRF.

Future high energy experiments \cite{Rowell2005} 
may be able to see features in the VHE spectra
of SNRs due to absorption on the infrared component of the ISRF.
Non-observation of absorption implies an upper bound on the 
the ISRF between the source and observer. 
Indirectly, this contrains the dust emission.


TAP is supported in part by the US Department of Energy,
IVM is supported in part by NASA APRA grant.

\bibliographystyle{aipproc}   


\end{document}